\documentclass[%
 reprint,
superscriptaddress,
 amsmath,amssymb,
pre,
]{revtex4-2}

\usepackage{graphicx}
\usepackage{xcolor}
\usepackage{dcolumn}
\usepackage{bm}
\usepackage{soul}

\newcommand{\me}{m_{\rm e}}

\newcommand{\wpp}{\omega_{\rm p}}

\newcommand{\Ma}{M_{\rm A}}

\begin{document}

\preprint{APS/123-QED}

\title{Relativistically Magnetized Collisionless Shocks in Pair Plasma\\
{\normalsize I. Solitons, Chaos, and Thermalization}}

\author{Arno Vanthieghem}
\email{arno.vanthieghem@obspm.fr}
\affiliation{Sorbonne Université, Observatoire de Paris, Université PSL, CNRS, LUX, F-75005 Paris, France}

\author{Amir Levinson}
\affiliation{School of Physics and Astronomy, Tel Aviv University, Tel Aviv 69978, Israel}

\date{\today}

\begin{abstract}
In this paper, the first in a series, we present a new theoretical model for the global structure and dissipation of relativistically magnetized collisionless shock waves. 
Quite remarkably, we find that in contrast to unmagnetized shocks, the leading energy dissipation channel does not involve collective plasma interactions. Rather, it is a consequence of nonlinear particle dynamics.
We demonstrate that the kinetic-scale shock transition can be modeled as a stationary system consisting of a large set of cold beams coupled through the magnetic field. The fundamental mechanism governing shock dissipation relies on the onset of chaos in orbital dynamics within quasiperiodic solitonic structures. We discuss the impact of upstream temperature and magnetization on the shock profile, recovering the magnetic field compression, downstream velocities, and heating expected from the Rankine-Hugoniot jump conditions. We deduce a rate of entropy generation from the spectrum of Lyapunov exponents and discuss the thermalization of the beam distribution. Our model provides a general framework to study magnetized collisionless shock structures.
\end{abstract}

\maketitle

\section{Introduction}

Relativistic magnetized shocks are widely believed to be the sources of the transient
radio and high-energy emissions observed in compact, strongly magnetized 
astrophysical systems, notably pulsars, magnetars and AGN jets.   These 
shocks might form due to the collision of a magnetized flow with an external medium, as in pulsar-driven nebulae and 
jet termination shocks, or due to nonlinear steepening of a compressive MHD wave, generated when the  magnetosphere 
of the compact object is disturbed, e.g., through star quakes~\cite{thompson1995, parfrey2013, Chen22, yuan22, mahlmann23}, 
collapse \cite{most2024monster}, or collisions with a compact companion~\cite{Hansen_2001, Most_2020, Beloborodov_2021}.

Early studies were primarily motivated by the detection of synchrotron emission in the Crab Nebula and other plerions~\cite{Hoshino_1992, Arons_1992, Gallant_1994}, as well as indications of relativistic motions in extragalactic jets.  The recent discovery of fast radio bursts (FRBs)~\cite{Lorimer_2007}, and the rapid advent of plasma kinetic simulations, initiated a tremendous resurgence of interest in relativistic magnetized shocks, particularly as sources of synchrotron maser emission~\cite{Lyubarsky_2014, Waxman_2017, Iwamoto_2019, Metzger_2019, Plotnikov_2019, Sironi_2021, Lyubarsky_2021, Babul_2020, Beloborodov_2023, Vanthieghem_2024b}.
FRBs can be generated as high-frequency precursor waves in monster shocks~\cite{Vanthieghem_2024b}, that form in the inner magnetosphere of a magnetar upon abrupt steepening of a millisecond fast magnetosonic wave~\cite{Beloborodov_2023}, or in shocks that form by the interaction of an escaping magnetosonic pulse with the surrounding medium well beyond the light cylinder ~\cite{Lyubarsky_2014, Waxman_2017, Metzger_2019, Lyubarsky_2021, Khangulyan_2022}.
Potential formation of monster shocks has been indicated recently also in global MHD simulations of delayed collapse of a magnetar~\cite{most2024monster}, as well as the interaction of neutron star magnetospheres in binary neutron star mergers \cite{Beloborodov_2023}, predicting early electromagnetic signals from these events. 

Much effort has been devoted to the development of a self-consistent, kinetic description of relativistic magnetized shock waves, using the Particle-In-Cell (PIC) simulation method in various geometries and compositions~\cite{Langdon_1988, Hoshino_1992, Gallant_1992, Gallant_1994, Iwamoto_2019, Plotnikov_2019, Sironi_2021, Vanthieghem_2024b}. A common picture has emerged from these numerical experiments for shocks propagating in highly magnetized plasmas, with magnetization $\sigma\,\gg\,0.1$, 
where $\sigma$ denotes the ratio of the Poynting flux to plasma momentum flux. 
In this regime, a strongly nonlinear and weakly dissipative structure forms at the foot of the shock transition layer. Identified as fast magnetosonic solitons, theoretical and numerical studies have characterized the physics of this leading nonlinear solitary wave~\cite{Kennel_1976, Alsop_1988, Gallant_1992, Gallant_1994} and discussed its 
role in the generation of a precursor wave, generally attributed to maser-synchrotron emission from unstable ring distributions formed in the soliton crossing region~\cite{Hoshino_1991}, albeit recent PIC simulations point towards more subtle emission mechanisms yet to be identified~\cite{Plotnikov_2019, Sironi_2021,Vanthieghem_2024b}. While extensive work exists on the analytical modeling of solitary structures, our understanding of the global shock transition and dissipation is purely defined by \emph{ab initio} kinetic simulations, and a theoretical description is still lacking.

This paper is the first in a series of papers presenting a new theory 
of relativistically magnetized collisionless shock waves propagating in a symmetric pair plasma. 
It focuses on the shock structure and dissipation mechanism, showing that dissipation occurs through the onset of chaos in the particle orbits. The analysis is carried out in stages; in Section~\ref{sec 1}, we employ a single fluid approach to derive a solution describing an infinite chain of periodic solitons, 
akin to the underlying microphysical shock structure (Sec.~\ref{sec 1.1}).
We then show (sec.~\ref{sec 1.2}) that particle trajectories in this solitary structure are unstable, leading to exponential divergence of 
orbits over a scale comparable to the distance between neighboring solitons. In section~\ref{sec 2}, we model the global transition using a self-consistent multifluid approach. We show that our model captures precisely the shock profile and jump conditions (Sec~\ref{sec 2.1}). To understand the global dynamics, we quantify the chaotic nature of the shock formation process, particularly the jump in entropy across the shock transition, and discuss the thermalization scale of the plasma (Sec~\ref{sec 2.2}). We conclude and summarize our results in Sec.~\ref{sec 3}.

\section{periodic Chain of Solitons} \label{sec 1}

Strongly magnetized shock waves feature a leading solitary structure at the interface between the shocked and the unshocked plasma in the regime of $\sigma \gg 0.1$. Extensive pieces of work have studied such highly nonlinear systems using reduced analytical approaches~\cite{Kennel_1976, Alsop_1988} and self-consistent numerical kinetic methods~\cite{Gallant_1992, Gallant_1994}.  Except for phenomenological models inferred from ab initio PIC simulations~\cite{Gallant_1994, Plotnikov_2018, Babul_2020, Sironi_2021}, the focus has mostly been on modelling them as independent structures. As a first step towards the characterization of the shock, this section aims at building a trail of stationary relativistic magnetosonic solitons. We will then show that the adopted approach can be generalized to capture the global shock structure. As of now, and for the sake of clarity, all quantities are expressed in the soliton rest frame, which is assumed to be stationary. 

\begin{figure}
  \centering
  \includegraphics[width=1.\columnwidth]{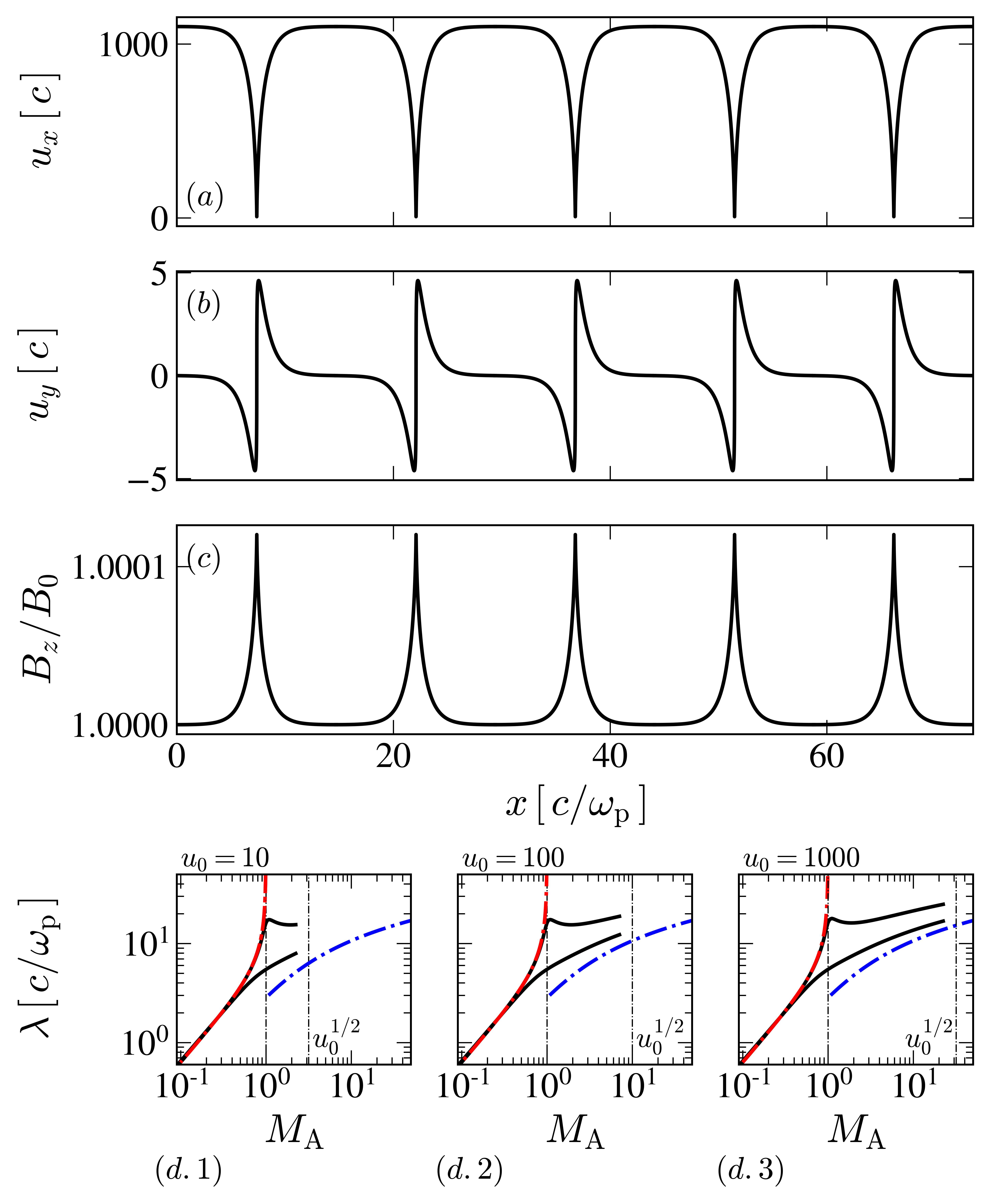}
  \caption{Solution to the periodic system of fast magnetosonic solitons. Panels (a)-(c), respectively, display the profiles of the longitudinal and transverse four-velocity and the magnetic field, for $\sigma\,=\,10^4$, $u_0\,=\,1000$, and $\delta \tilde{u}_x(0) = 0.1$. The resulting period is $\lambda\,\simeq\,14.7\,c/\wpp$. Panel (d.1-3)  depict the range of wavelengths of the soliton train for $u_0\,=\,10^1,\,10^2,\,10^3$, as indicated, and for $\delta \tilde{u}_x(0) = 0.005$ (upper curve) and $0.3$ (lower curve). The rightmost black vertical dot-dashed line in each panel corresponds to the approximate validity limit of the single-fluid approach, $\gamma_{\rm sh} = \sigma$. The red line corresponds to the exact wavelength in the linear regime, $\Ma\,\ll\,1$, and blue dot-dashed line to the wavelength estimated from the linearized solution for $\Ma\,\gtrsim\,1$, given by  Eq.~\eqref{eq:lambda}.}
  \label{fig:period}
\end{figure}

\subsection{Stationary State}\label{sec 1.1}

In what follows, we restrict our analysis to relativistically magnetized shocks where $\sigma \gg u_{\rm sh} \sim \gamma_{\rm sh}$, with $\gamma_{\rm sh}^2\,=\,1+u_{\rm sh}^2$ where $u_{\rm sh}$ is the shock four-velocity in the upstream frame. In this regime, the longitudinal bulk velocity of the flow is sign-preserving over the soliton crossing.
Here, we arbitrarily assume the flow to drift along $+\hat{\bm{x}}$ ($u_{x} >0$). We further assume that the magnetic field is perpendicular to the flow along the $z$-direction. By symmetry of electron and positron trajectories, 
the whole system can then be captured by a single fluid model as long as $\sigma \gg \gamma_{\rm sh}$~\cite{Alsop_1988}. The single fluid limit therefore corresponds to $\sigma\,\sim\,\gamma_{\rm sh}$ or, equivalently, $\Ma\,\sim\,\sqrt{\gamma_{\rm sh}}$ where we introduced the Alfvénic Mach number, $\Ma\,=\,|u_{\rm sh}|/\sqrt{\sigma}$.
We first recall the system of equations derived from the conservation of particle number, energy and momentum,
coupled to Amp\`ere-Maxwell equations:
\begin{align}
    &\partial_{\tilde{x}} \tilde{u}_x \,=\, \frac{\tilde{B}_z}{\Ma}\,\frac{\tilde{u}_y}{\tilde{u}_x} \,,\label{eq:1_fluid_1}\\
    &\partial_{\tilde{x}} \tilde{u}_y \,=\, - \frac{\tilde{B}_z}{\Ma} +  \frac{1}{\Ma} \frac{\gamma}{\tilde{u}_x} \,,\label{eq:1_fluid_2}\\
    &\partial_{\tilde{x}} \tilde{B}_z \,=\, -\Ma \frac{\tilde{u}_y}{\tilde{u}_x}  \,,\label{eq:1_fluid_3} \\
    &\partial_{\tilde{x}} \tilde{E}_y \,=\, 0  \,,\label{eq:1_fluid_4} 
\end{align}
where the electric and magnetic fields are normalized to the upstream proper magnetic field, $\tilde{B}_z\,=\,\gamma_0\,B_z/B_0$ and $\tilde{E}_y\,=\,\gamma_0\,E_y/B_0$, $B_0$ being the apparent field in the soliton frame, and velocities are normalized to the upstream velocity $u_{\rm sh}\,=\,u_0$, as measured in the soliton frame, $\tilde{u}_x\,=\,u_x/u_0$ and $\tilde{u}_y\,=\,u_x/u_0$. Equation \eqref{eq:1_fluid_4} readily implies that the electric field is constant throughout, $E_y = B_0 \beta_0$ ($\tilde{E} = u_0$), 
where $\beta_0 \,=\, u_0\,(1+u_0^2)^{-1/2}$ is the corresponding three velocity. To obtain a non-trivial solution, we introduce a small perturbation 
in the initial upstream velocity, viz., $u_x(0) = u_0 + \delta u_x(0)\,=\,u_0\, \tilde{u}_x(0)\,=\,u_0\,[1+\delta \tilde{u}_x(0)]$. As will be clarified later, we normalize distances $x$ to the proper upstream plasma skin depth, $c/\wpp$, where $\wpp\,=\,\sqrt{4 \pi (n_+ + n_-) e^2/m_{\rm e}}$ is the plasma frequency, such that ${\tilde{x}}\,=\,\wpp x/c$.  Here $n_+$ and $n_-$ are, respectively, the number density of the positrons and electrons upstream of the shock.
The above set of equations can be readily reduced to a second-order differential equation in $B_z$, equivalent to the decoupled equations in~\cite{Kennel_1976}.

We seek periodic solutions of the above system of coupled ODEs in the strongly nonlinear regime, where there is no general analytical solution. We treat this problem as a boundary value problem. An analytical solution is first obtained by linearizing equations \eqref{eq:1_fluid_1}-\eqref{eq:1_fluid_3} 
in the regime of subcritical fast magnetosonic flow, $M_A\,=\,|u_0|/\sqrt{\sigma}\,\ll\,1$, irrelevant to shocks.
We obtain the functional form as well as the wavenumber $ck/\wpp\,=\,2\pi \lambda^{-1}\,c/\wpp\,=\,\sqrt{\sigma/u_0^2 - 1}$. The scaling of the wavelength $\lambda$ with $\Ma$ 
in the linear regime is depicted by the red line in Fig.~\ref{fig:period} panels d.1-3 for comparison with the nonlinear case, discussed below. 
The train of nonlinear periodic single-fluid solution is then obtained iteratively from this initial analytical guess.

The typical profiles of the nonlinear structure are shown in the three top panels of Fig.~\ref{fig:period}. Successive compressions of the magnetic field are associated with Larmor gyration of the fluid in the changing magnetic field, resulting in formation of localized current filaments (the peaks in $u_y$). A comprehensive account of the general properties of a single magnetosonic soliton can be found in~\cite{Alsop_1988}. Our formalism extends it to a soliton chain, as seen in recent kinetic simulations~\cite{Plotnikov_2018,Sironi_2021,Vanthieghem_2024b}.  
One salient feature arising from these simulations is a quasi-stationary distance between consecutive solitons. This property is nicely reproduced by our solutions,
as seen in the example shown in Fig.~\ref{fig:period}($a-c$).  Below, we derive an analytic estimate of this distance that elucidates its scaling.

The scaling of the wavelength of the soliton train with $M_A$ in the linear and nonlinear regimes is exhibited in  
Fig.~\ref{fig:period}(d.1-2), for the choice of 4-velocities $u_0\,=\,10,\,100,\,1000$, different magnetizations, $\sigma_{u_0=10}\,=\,[10^1,\,10^2]$, $\sigma_{u_0=100}\,=\,[10^2,\,10^4]$, and $\sigma_{u_0=1000}\,=\,[10^3,\,10^8]$, and  various initial velocity perturbations in the range $\delta \tilde{u}_x(0)\,\in\,[0.005,\,0.3]$. In each case, we observe the expected relation, $ck/\wpp\,=\,( \Ma^{-2} - 1 )^{1/2}$, for $M_{\rm A}\,<\,1$, while for $M_{\rm A}\,>\,1$, the period exhibits a weak dependence on the Alfv\'enic Mach number, remaining close to $ck/\wpp\,\simeq\,1$. To understand this property, we note that in the nonlinear regime, the trough of the wave dominates the wavelength;  the size of the crest is proportional to the Larmor radius, which for large $\sigma$ is much smaller than the skin depth. In the trough, we have $\tilde{B}_z \,\simeq\, \gamma_0  ( 1 + \delta \tilde{B}_z )$, with $\delta \tilde{B}_z\,\ll\,1$ also in the strongly nonlinear regime. Hence, we can linearize the system around the trough, assumed to be located at $x\,=\,0$. From the linearized system, with the initial conditions $\delta \tilde{B}_z(0) \,=\, 0$ and $\partial_{\tilde{x}} \delta \tilde{B}_z(0) \,=\, 0$, we obtain an analytical solution for the magnetic field profile:
\begin{align} \label{eq:est_dBz}
    \delta \tilde{B}_z\,=\,\frac{\Ma^2}{\Ma^2 - 1}\,\frac{\delta \tilde{u}_x(0)}{\gamma_{0}^{2}} \left[ \cosh\left( \frac{\sqrt{\Ma^2 - 1}}{\Ma} \tilde{x} \right) - 1 \right]\,.
\end{align}
Equation~\eqref{eq:est_dBz} is only valid far from the solitons peaks where changes in velocities and magnetic field become nonlinear. An analogous form of the above expression was found in~\cite{Alsop_1988} [see Eq. (32) there]. The characteristic distance between neighboring solitons is derived from
Eq. \eqref{eq:est_dBz}, and the condition $\tilde{u}_x(x)\,=\,1$ constrained by the first integral $\partial_{\tilde{x}} ( \tilde{u}_x \,+\, \tfrac{1}{2} \tilde{B}_z^2/\Ma^2 )\,=\,0$. This yields the relation $\Ma^2\,=\,\cosh( \sqrt{ 1 -\Ma^{-2}}\,\tilde{x} )$, with the argument $\tilde{x}\,\simeq\,\tfrac{1}{2}\lambda\,\wpp/c$, from which one obtains  the following analytic approximation for the wavelength:
\begin{align}\label{eq:lambda}
    \lambda\,\simeq\, 2\, \frac{\ln \left( \Ma^2 + \sqrt{\Ma^4 - 1} \right)}{\sqrt{ 1 - \Ma^{-2} }} \, c/\wpp \,,
\end{align}
which exhibits a logarithmic scaling in $\Ma$, in good agreement with the exact solution, shown by the blue curves in Fig.~\ref{fig:period}(d.1-3). The scaling of the wavelength is determined in the trough of the wave. In this region, the single fluid approximation remains valid, even in the regime of reflected particles~\cite{Alsop_1988}. Consequently, we expect that Eq.~\eqref{eq:lambda} holds true for a wide range of soliton-mediated shock waves.

\begin{figure}
  \centering
  \includegraphics[width=1.\columnwidth]{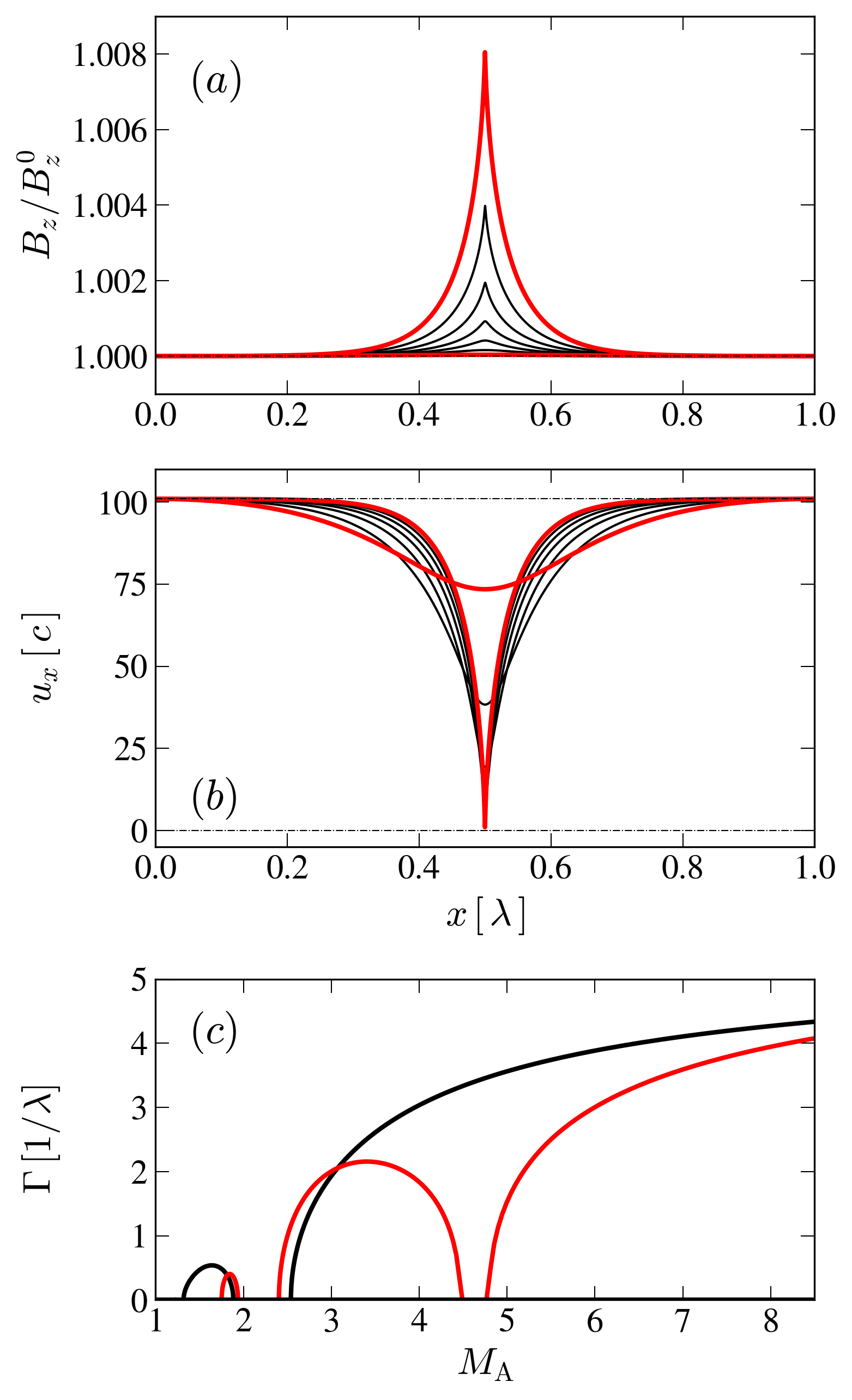}
  \caption{Unperturbed structure of a single soliton in the chain, and dependence of the growth rate of trajectories perturbations on $\Ma$. The two panels (a) and (b) show, respectively, the magnetic field compression and longitudinal flow velocity over one period, for $u_0\,=\,100$, $\delta \tilde{u}_x(0)\,=\,0.01$, and $\sigma\,=\,125,\,250,\,500,\,1000,\,2000,\,4000,\,8000$ with extrema highlighted by the red lines. The bottom panel (c) shows the Floquet characteristic exponents as 
  a function of $M_A$, for $u_0\,=\, 100$, $\delta \tilde{u}_x(0)\,=\,0.01$ (red) and $\delta \tilde{u}_x(0)\,=\,0.1$ (black). Positive exponents indicate exponential 
  divergence of the test-particle orbits.}
  \label{fig:chaos}
\end{figure}

\subsection{Onset of Chaotic Orbits}\label{sec 1.2}

In the previous section, we explored the properties of a stationary, periodic solitary structure, generated by a single fluid beam in a supercritical fast 
magnetosonic flow.  We showed that the degree of nonlinearity and the characteristic wavelength are controlled solely by the  Alfv\'en Mach number $\Ma$.
This elementary solitary structure can be generalized to describe a shock transition upon the inclusion of dissipative processes, e.g., plasma instabilities 
or chaotic orbital dynamics.  In what follows, we focus on identifying the dominant dissipation mechanism, which we propose is phase mixing among chaotic trajectories.
In a follow-up paper, we shall study the generation of plasma instabilities, which are subdominant, in an attempt to show that they can account for the generation of a high-frequency precursor wave, as seen in recent PIC simulations~\cite{Iwamoto_2019,Plotnikov_2019,Vanthieghem_2024b}.

As a preliminary step, we analyze, in this section, the stability of test particle trajectories interacting with the periodic multi-soliton structure
computed in Sec.~\ref{sec 1.1} (that is, we ignore any feedback on the system).  Our goal is to illustrate how chaos can emerge through the shock transition layer as particles repeatedly cross solitons. We intentionally omit feedback effects on the magnetic field at this point, providing only a typical scale for the divergence of trajectories prior to phase mixing. A self-consistent treatment of phase mixing in a quasi-isotropized system will be given in the next section. Here, we concentrate on characterizing small deviations from the initial equilibrium state.

In the limit of weak perturbations, the stability of the trajectories is obtained by linearizing the system of equations around the single-fluid solution. For each velocity component, we therefore have $\tilde{u}_i\,\equiv\,\tilde{u}_i(x) + \delta\tilde{u}^{(2)}_i(x) $, where the superscript $(2)$ designates trajectory perturbations, to distinguish from the initial variation used above to compute the stationary structure. To leading order in the perturbation, we obtain: 
\begin{align} \label{eq:Mathieu}
    \partial_{\tilde{x}} \delta \tilde{u}^{(2)}_x &\,=\, - \frac{\tilde{u}_y}{\tilde{u}_x^2} \frac{\tilde{B}_z}{\Ma} \,\delta \tilde{u}^{(2)}_x + \frac{\tilde{B}_z}{\Ma \tilde{u}_x}\,\delta \tilde{u}^{(2)}_y \,,\\
    \partial_{\tilde{x}} \delta \tilde{u}^{(2)}_y &\,=\, - \frac{1 + u_{0}^{2} \tilde{u}_y^2}{\Ma 
    \tilde{u}_x^2 \gamma} \,\delta \tilde{u}^{(2)}_x + \frac{u_0^2}{\Ma} \,\frac{\tilde{u}_y}{\tilde{u}_x \gamma} \, \delta \tilde{u}^{(2)}_y\,.
\end{align}
Given the periodicity of the coefficients, we use the fundamental theorem of the Floquet theory, from which we extract the maximum Floquet exponent, $\Gamma$, corresponding to the growth rate of the most unstable mode~\cite{Vanthieghem_2018}. 
The upper panels in Figure~\ref{fig:chaos} display the nonlinear profiles of the magnetic field (Fig.~\ref{fig:chaos}a) and longitudinal 4-velocity (Fig.~\ref{fig:chaos}b) 
of a single soliton in the periodic chain,
for $u_0\,=\,100$, $\delta \tilde{u}_x(0)\,=\,0.01$, and Mach number ranging from $\Ma\,=\,1.11$ ($\sigma\,=\,8000$) to $\Ma\,=\,8.9$ ($\sigma\,=\,125$). 
Fig 1c exhibits the dependence of the growth rate of perturbed test-particle orbits on $M_A$, for two cases, $\delta \tilde{u}_x(0)\,=\,0.01$ and $0.1$.
We observe regions of stability at low Mach numbers between $\Ma\,\sim\,1$ and $\Ma\,\sim\,4.7$. These features are characteristic of a Mathieu-type equation,
and are artifacts of our neglect of feedback. 

\begin{figure}
  \centering
  \includegraphics[width=1.\columnwidth]{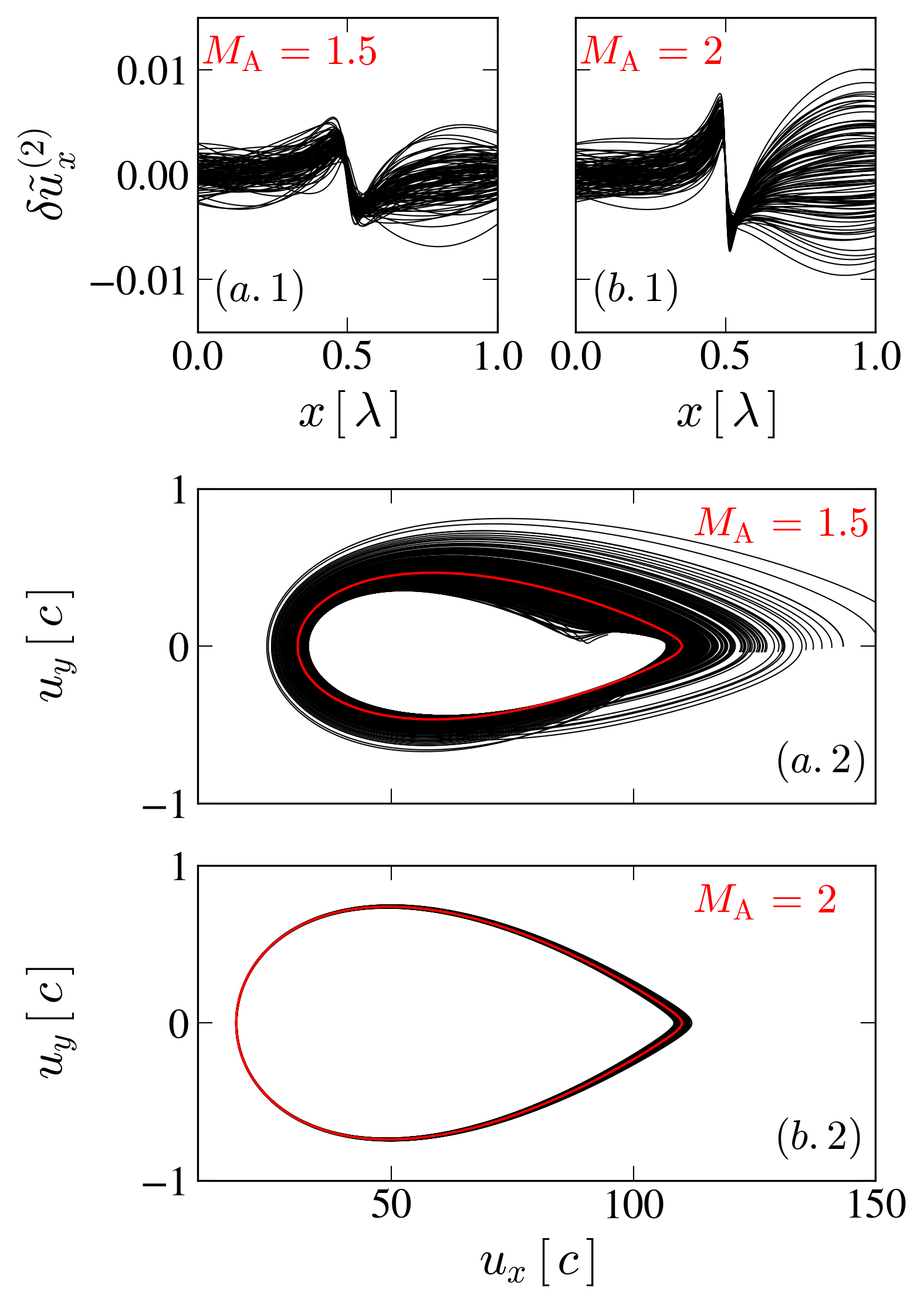}
  \caption{Orbits of 100 test-particles with initial spread drawn from a thermal distribution with a proper temperature $k_{\rm B}T\,=\,10^{-6}\,\me c^2$, computed for a single-fluid multi-soliton solution with $u_0\,=\,100$, $\delta \tilde{u}_x(0)\,=\,0.1$,  $\Ma\,=\,1.5$ (a) and $\Ma\,=\,2$ (b). Top (a-b.1): Perturbative orbits 
  $\delta \tilde{u}_x^{(2)}$ over the first crossing. 
  Bottom (a-b.2): Comparison of the phase-space profile over 10 consecutive solitons crossing.
  While a large velocity spread is observed in (b.1) for $\Ma\,=\,2$, the global test-particle orbit is stable (b.2); As predicted in Fig.~\ref{fig:chaos}(c), the test-particle orbits are unstable for $\Ma\,=\,1.5$ and become stable for $\Ma\,=\,2$. The red lines in (a-b.2) delineate the stationary solution.}
  \label{fig:orbit}
\end{figure}

Figure~\ref{fig:orbit} exhibits the behavior of particle dynamics for two precomputed single-fluid solitary structures, with $u_0=100$, $\delta \tilde{u}_x(0)\,=\,0.1$
and $\Ma\,=\,1.5$ [panel (a.2)],  $\Ma\,=\,2$ [Panel (b.2)].  It confirms the behavior anticipated from the Floquet analysis (Fig. \ref{fig:chaos}c), specifically,
unstable orbits for $M_A=1.5$ and stable orbits for $M_A=2$.
The trajectories have been obtained through direct integration on the interpolated magnetic field using the Vay pusher algorithm~\cite{Vay_2008}. Panels~\ref{fig:orbit}(a-b.1) indicate clear evidence of momentum spread, even in the stable case, over a single soliton crossing. This aligns with the typical values of the Floquet characteristic exponent shown in Fig~\ref{fig:chaos}(c), where the e-folding distance is comparable to the wavelength. The stability regions mentioned earlier are symptomatic of the test particle integration (i.e., neglect of feedback). The present approach successfully fulfils the initial intent of showing the rapid divergence of trajectories over one soliton crossing. However, it is crucial to properly include the particle feedback on the magnetic field to accurately describe the global stability, which will reveal the erasure of these zones of marginal stability~\footnote{We refer to marginal stability here, as stability is ensured for only very small initial perturbations around equilibrium.}.

\begin{figure*}[t]
  \centering
  \includegraphics[width=1.\textwidth]{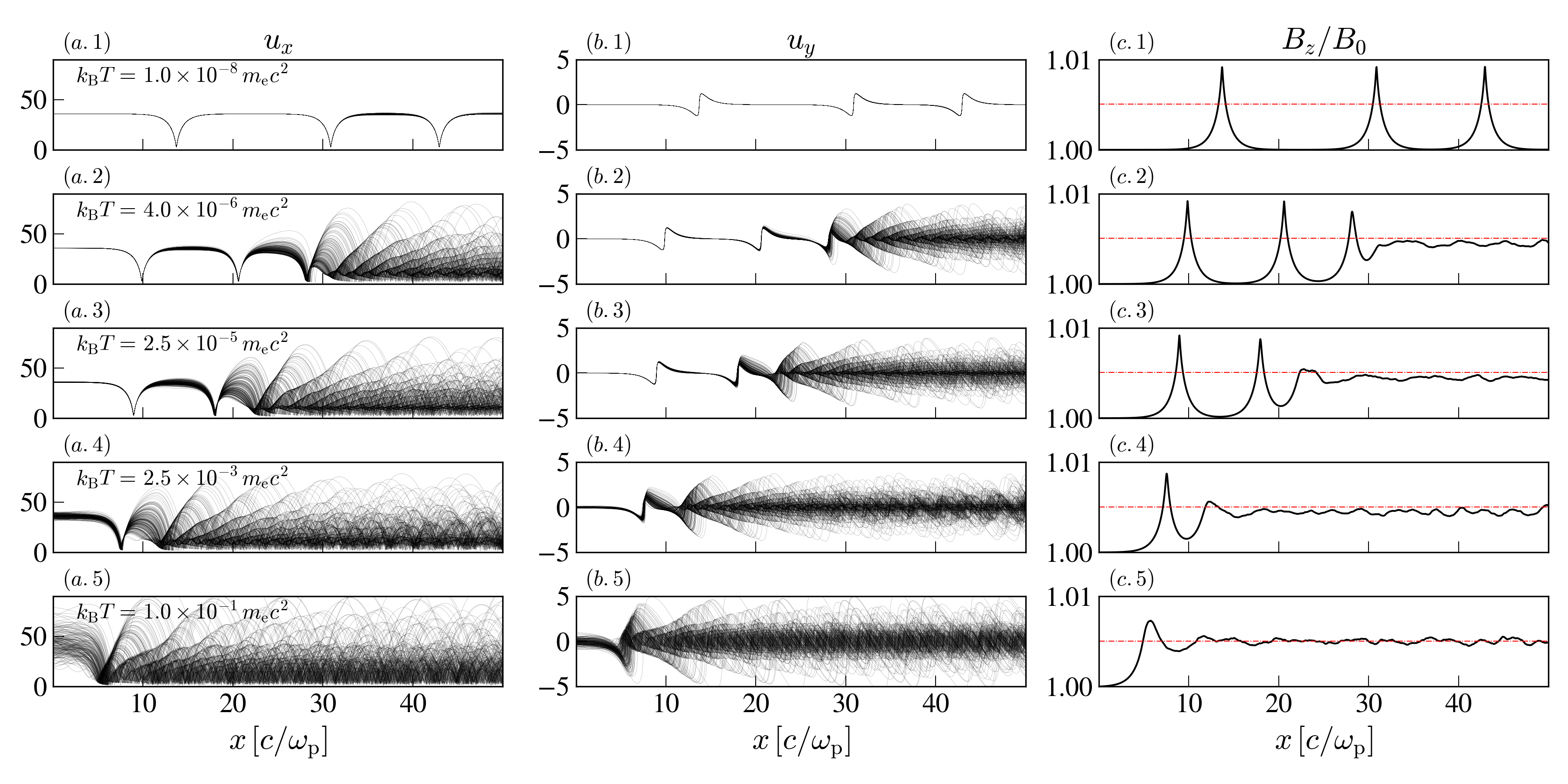}
  \caption{Synthetic shock profiles for $N\,=\,200$,  $\sigma\,=\,100$, $u_0\,=\,35.8$ and, from top to bottom, various temperatures $k_{\rm B} T_{\rm e}\,=\,10^{-8},\,4\times 10^{-6}, 2.5\times10^{-5}, 2.5\times10^{-3}, 10^{-1}\,m_{\rm e} c^2$. From left to right, the panels correspond, respectively, to the longitudinal and transverse four velocities and the magnetic field compression ratio, measured in the shock-front (soliton) frame. The horizontal red dashed lines in the rightmost panels mark the expected magnetic field compression ratio obtained from the Rankine-Hugoniot jump conditions.}
  \label{fig:Var_T}
\end{figure*}

\begin{figure*}[t]
  \centering
  \includegraphics[width=1.\textwidth]{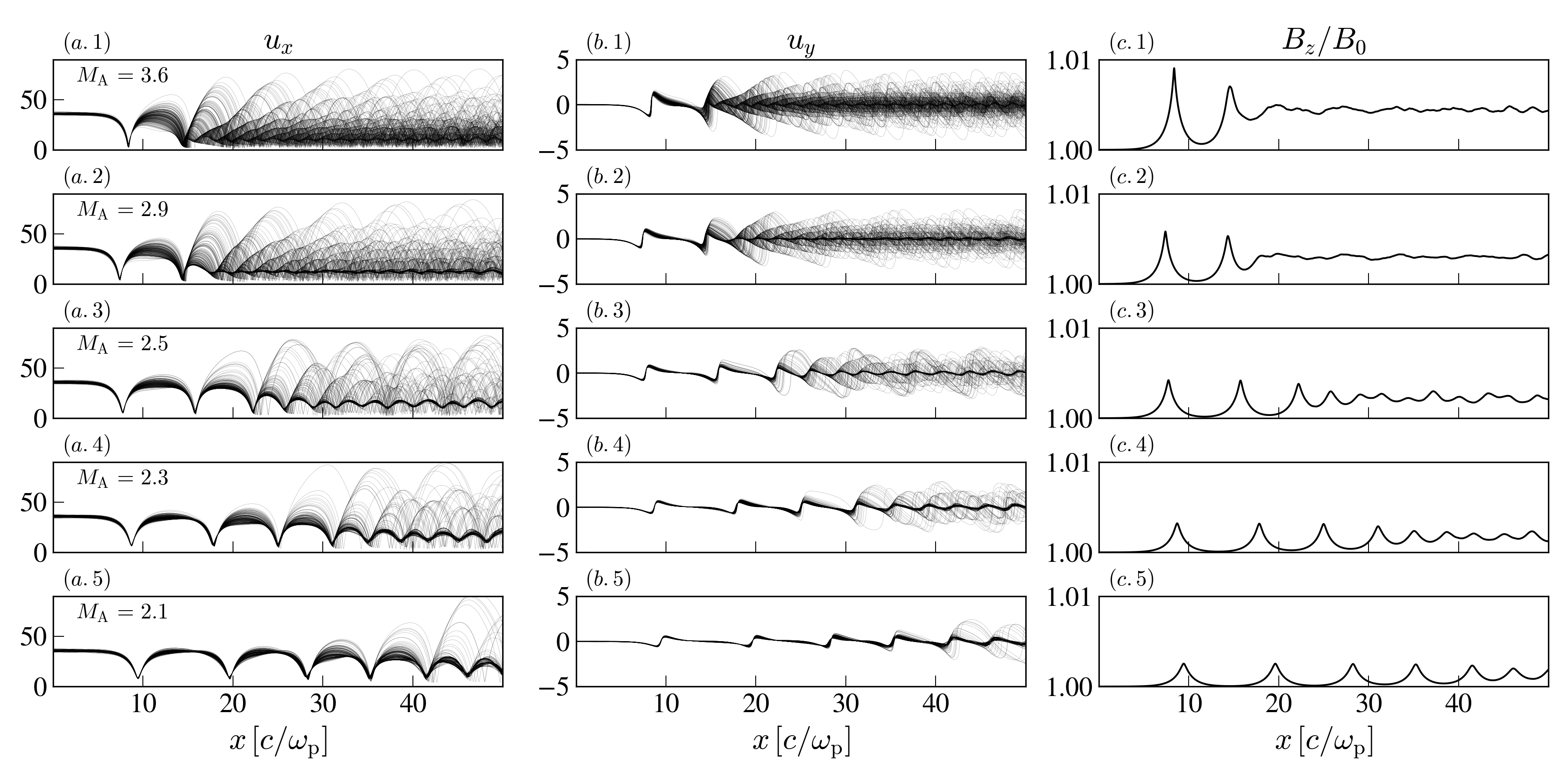}
  \caption{Same as Fig.~\ref{fig:Var_T} for $u_0\,=\,35.8$, $k_{\rm B} T_{\rm e}\,=\,4\times10^{-4}\,m_{\rm e} c^2$, and various values of $\sigma$ corresponding,
  from top to bottom, to $\Ma\,=\,3.6,\,2.9,\,2.5,\,2.3,\,2.1$.}
  \label{fig:Var_Ma}
\end{figure*}

\section{Shock Structure and Dissipation}\label{sec 2}

In the previous section, it was demonstrated that test particle motion in a stationary, multi-soliton chain becomes stochastic following 
a series of soliton crossings.  However, the feedback of the growing trajectory spread on the solitary structure was neglected, lacking
the essential coupling required to establish a complete, self-consistent shock transition.  In this section, we incorporate the feedback
of the particle system on the magnetic field. We show that this is sufficient to account for shock dissipation and downstream thermalization and that the fully coupled system indeed describes a complete, stationary shock transition.

It has been shown that thermal effects are marginally captured in the limit of a large number of coupled cold fluids~\cite{Dawson_1960}.  Here, we use a similar approach to approximate kinetic effects at play in the shock transition up to an arbitrary order of precision.

\subsection{The Large N-Fluid Limit}\label{sec 2.1}

We model the symmetric pair plasma as a collection of $N$ neutral fluids, where each neutral fluid consists of 2 beams, one of electrons and one of positrons.
Since the plasma is symmetric, it is sufficient to solve the equations only for one species; henceforth, we use the positron beams as our
variables, whereby for the $i$-th fluid we have $u^i_{x}\,=\,u^i_{e,x}\,=\,u^i_{p,x}$ and $u^i_{y}\,=\,-u^i_{e,y}\,=\,u^i_{p,y}$.
Each fluid beam thus obeys Eq.~\eqref{eq:1_fluid_1}-\eqref{eq:1_fluid_2}, but the net electric current density is the sum over all contributions:
$j_y = e\sum_i (n_{p}^i u^i_{p,y} - n_{e}^i u^i_{e,y}) = 2e\sum_i n^iu_y^i = \frac{2e n_0u_0}{N}\sum_i(u^i_y/u^i_x) $.  With the normalization adopted above, Amp\`ere's law reads:
\begin{align}\label{eq:B_Nfluid}
    \partial_{\tilde{x}} \tilde{B_z} \,=\, - \Ma\,\frac{1}{N}\,\sum^N_i\,\frac{\tilde{u}^i_y}{\tilde{u}^i_x}\,,
\end{align}
where the index $i$ runs over all of the $N$ fluids. Each fluid is initialized at $x\,=\,0$ according to a J\"uttner-Synge distribution, thus mimicking the initial kinetic or fluid thermal spread. To be concrete, the initial conditions for each beam are: $n^i(0)u^i_x(0)\,=\,j_u/N$, $u^i_x(0)\,=\,u_u + \delta u^i_x$, $u^i_y(0)\,=\,\delta u^i_y$, here
$j_u = n_u u_u$, with $n_u$ and $u_u$ being the density and 4-velocity of the upstream flow, respectively, $\sum_i n^i(0) \delta u^i_x\,=\,0$, and the initial variations, $\delta u^i_x$ and $\delta u^i_y$, are drawn from a 2D J\"uttner-Synge distribution. In the cold temperature limit, we recover the single-fluid solution. While an initial relative drift was necessary to trigger the soliton formation in the previous section, the effective thermal noise is sufficient here. 

From Sec.~\ref{sec 1.2}, we anticipate a strong dependence of the shock profile on the initial thermal spread due to the exponential divergence of trajectories. The striking similarity between our model and kinetic simulations discussed in the literature (see~\cite{Plotnikov_2019}, for instance) is blatant in Fig.~\ref{fig:Var_T}, where solutions obtained for $N = 200$, $\Ma = 3.58$, $\sigma = 100$, and various initial thermal spreads in the range $k_{\rm B} T \in [10^{-8}, 10^{-1}] \me c^2$ are shown. We observe a decrease in the cavity size at each crossing, corresponding to an increased spread in the distribution of fluid velocities. The thermal spread is analogous to the initial velocity perturbation $\delta \tilde{u}_x(0)$ in the one-fluid model, which is consistent with Fig.~\ref{fig:period}(d.1-3), where the period decreases with increasing $\delta \tilde{u}_x(0)$. A few soliton crossings quickly redistribute part of the directed kinetic energy into the internal energy of the distribution of longitudinal and transverse velocities [see panels (a.1-5) and (b.1-5) of Fig.~\ref{fig:Var_T}, respectively]. As discussed further below, we find a very good agreement between the downstream compressed magnetic field obtained from our model and that derived from the Rankine-Hugoniot jump conditions (red lines in panels c.1-5). 
In agreement with~\cite{Babul_2020}, an initial mildly relativistic thermal spread on the order of $k_{\rm B} T = 10^{-1} \me c^2$ disrupts the cavity. The effect of the shock structure on the radiative efficiency in terms of upstream thermal spread goes beyond the scope of the present study and will be discussed in the second paper of this series.

The Alfv\'enic Mach number $\Ma$ dictates the shock dynamics. As discussed in Section~\ref{sec 1.2}, the Floquet characteristic exponents vanish as $\Ma \to 1^+$. This implies that for the same initial velocity perturbation, we expect to observe more generations of solitons as $\Ma$ decreases (or $\sigma$ increases for a constant upstream velocity). Figure~\ref{fig:Var_Ma} illustrates this behavior, showing the evolution of the shock profile for decreasing values of $\Ma$ (from top to bottom). As the system becomes less nonlinear, more soliton crossings are required for the onset of chaotic orbits. This is evident in Fig.~\ref{fig:Var_Ma}, where for a given temperature, $k_{\rm B} T_{\rm e} = 4 \times 10^{-4} \, m_{\rm e} c^2$ in this example, a single cavity (i.e., two solitons) is observed for $\Ma = 3.6$, whereas at least six cavities are visible for $\Ma = 2.1$. This trend underscores the critical role of the Alfv\'enic Mach number in the formation and evolution of solitons. Furthermore, as $\Ma$ decreases, the shock transition becomes more gradual, requiring multiple soliton interactions to redistribute the energy effectively. The strong dependence of the shock profile on the upstream temperature motivates the intricate interplay between nonlinear effects, such as the coupling between associated upstream emission, the upstream plasma, and the soliton dynamics. 

\begin{figure}[t]
  \centering
  \includegraphics[width=1.\columnwidth]{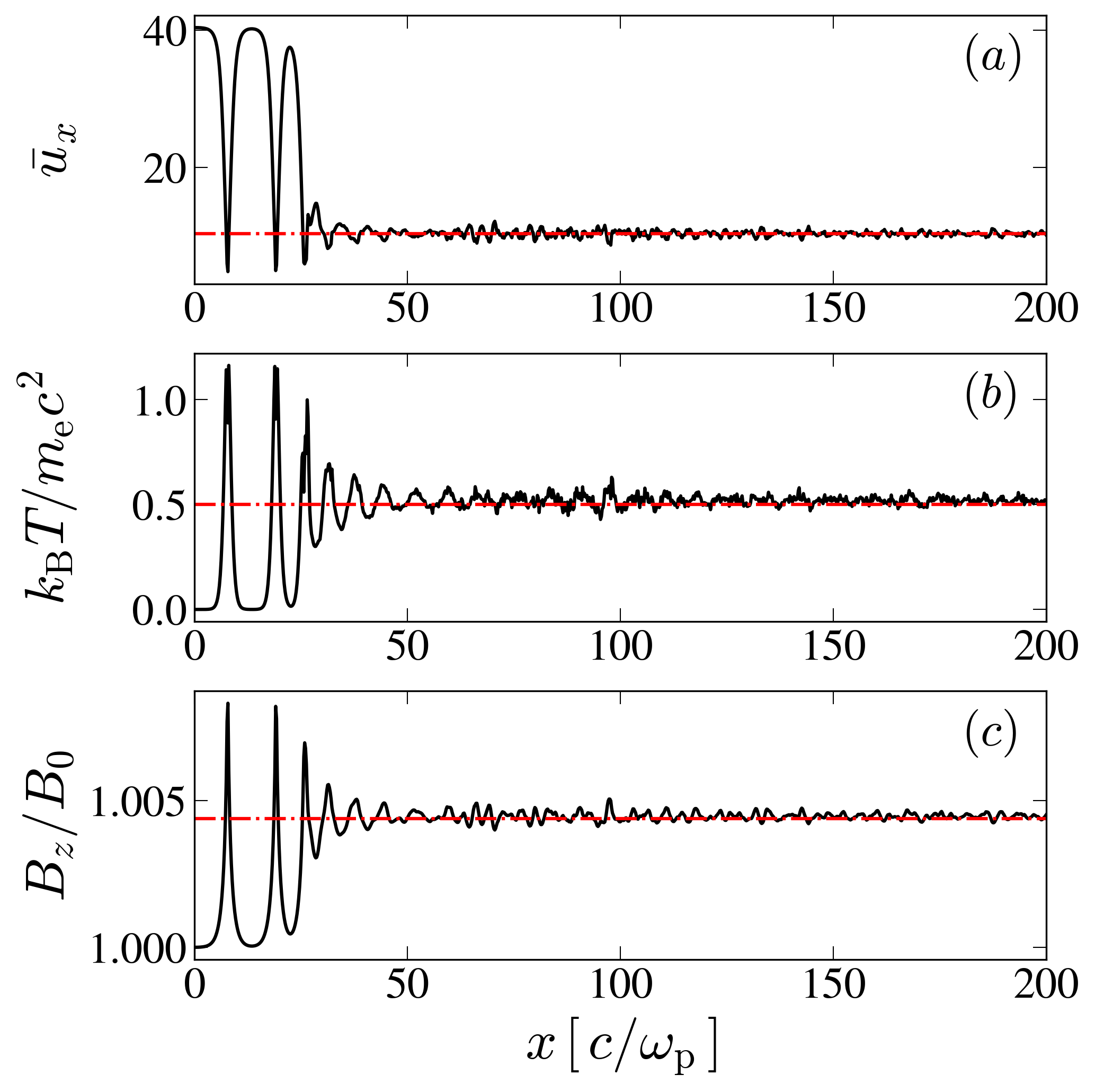}
  \caption{Comparison with the Rankine-Hugoniot jump conditions for $N\,=\,10^4$, $\Ma\,=\,40$ and $\sigma\,=\,100$. From top to bottom, the panels show (a) the longitudinal bulk-flow velocity, (b) the effective transverse temperature profile, (c) the compression factor across the shock transition. The red dot-dashed lines correspond to the exact jump conditions for the same parameters.}
  \label{fig:RH}
\end{figure}

To assess the validity of our approach, it is essential to perform a direct comparison with the Rankine-Hugoniot jump conditions. Such a comparison requires defining the moments of the velocity distribution of cold fluids as a discretized limit of the kinetic distribution. The relevant moments of the distributions are:
\begin{align}
    \bar{\beta}_x &= \bar{u}_x/\sqrt{1+\bar{u}_x^2} = \frac{1}{N}\sum_{i=1}^N \frac{u_x^i}{\gamma^i} \,, \label{eq:bar_ux} \\
    T &\equiv T_\perp = \frac{\me c^2}{k_{\rm B} N}\sum_{i=1}^N \frac{u_y^i u_y^i}{\gamma^i} \,.\label{eq:Temp}
\end{align}
Together with the compression ratio, we compare the evolution of these moments. For regimes of extreme magnetization where the downstream temperature becomes mildly relativistic~\cite{Kennel_1984}, using the relativistic limit of the adiabatic index in two dimensions ($\Gamma_{\rm ad} = 3/2$) leads to incorrect jump conditions. As shown in~\cite{Vanthieghem_2019}, the adiabatic index of a Maxwell-J\"uttner distribution in arbitrary dimensions obeys:
\begin{align}
    \Gamma_{\rm ad} &= \frac{K_{\frac{d+3}{2}}(\mu) - K_{\frac{d+1}{2}}(\mu)}{K_{\frac{d+3}{2}}(\mu) - \left(1 + 1/\mu\right)K_{\frac{d+1}{2}}(\mu)} \,, \\
    &= 2 - \frac{1}{2 + \mu} \qquad (d = 2) \,,
\end{align}
where $K_\alpha$ is the $\alpha^{\rm th}$-order modified Bessel function of the second kind, $d$ is the phase space dimension, and $\mu = \me c^2 / (k_{\rm B} T)$. Since the particle motion is purely transverse to the magnetic field, the phase space dimension is $d\,=\,2$. Figure~\ref{fig:RH} shows the spatial profile of the effective bulk flow velocity defined in Eq.~\eqref{eq:bar_ux}, the effective temperature defined in Eq.~\eqref{eq:Temp}, and the compression ratio of the magnetic field for profiles corresponding to $10^4$ plasma beams for $u_0\,=\,40$, $\sigma\,=\,100$, $\Ma\,=\,4$. Each of these quantities is found to be in very good agreement with the exact value of the jump conditions computed with the temperature-dependent adiabatic index $\Gamma_{\rm ad}$. We note that the amplitude of the fluctuations seen around the analytic jump conditions decreases as the number of fluids increases.

\begin{figure}[t]
  \centering
  \includegraphics[width=1.\columnwidth]{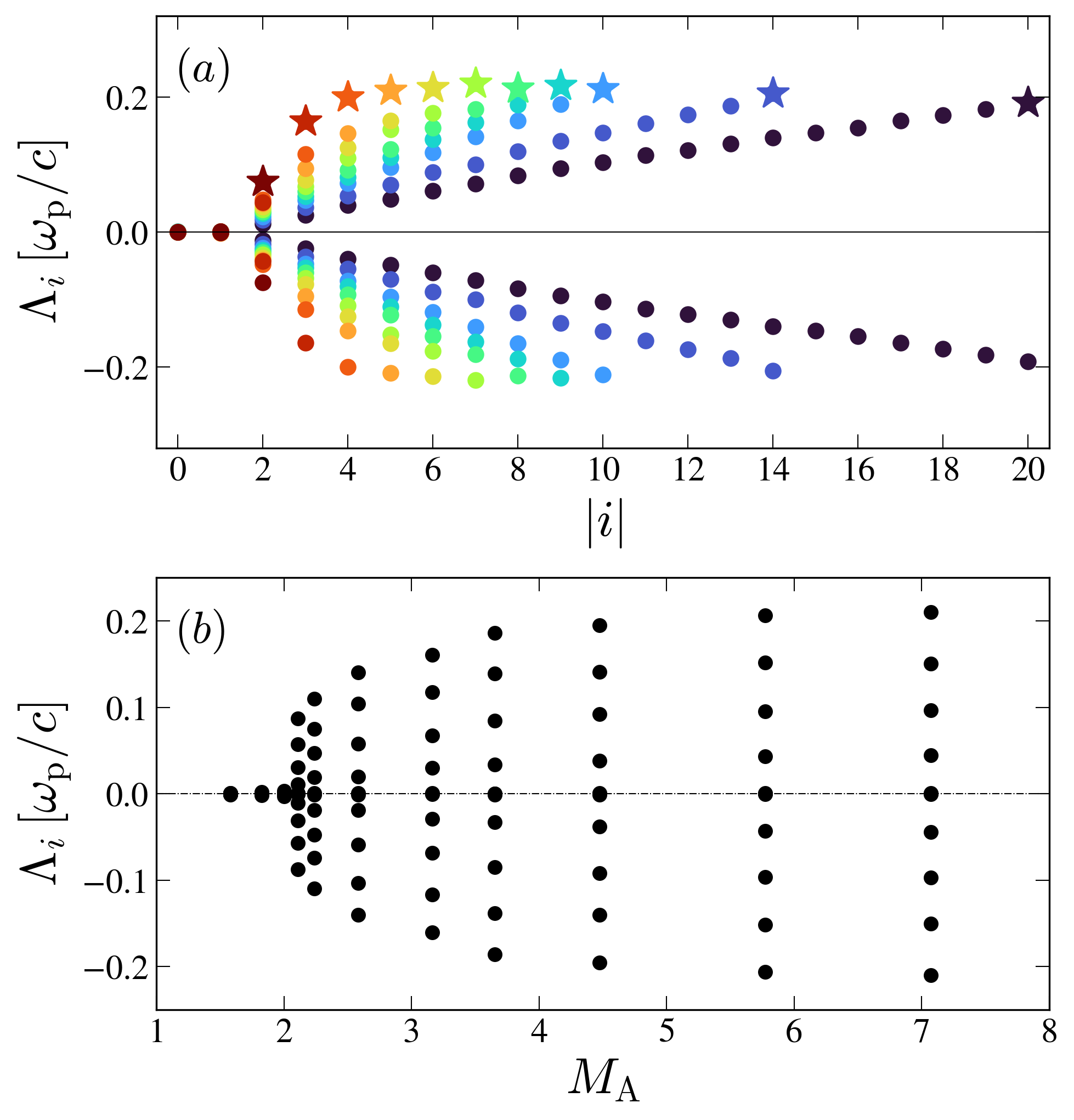}
  \caption{(a) Spectrum of the Lyapunov exponents $\Lambda_i$ for an increasing number $N$ of fluids for $u_0\,=\,100$ and $\sigma\,=\,200$. For each $N$ the maximum Lyapunov exponent corresponds to the starred value. (b) Spectrum of the Lyapunov exponents as a function of $\Ma$ for $N\,=\,5$, $u_0\,=\,100$ and $\sigma\,=\,[200,\,8000]$. }
  \label{fig:Lyapunov}
\end{figure}

\begin{figure*}[t]
  \centering
  \includegraphics[width=1.\textwidth]{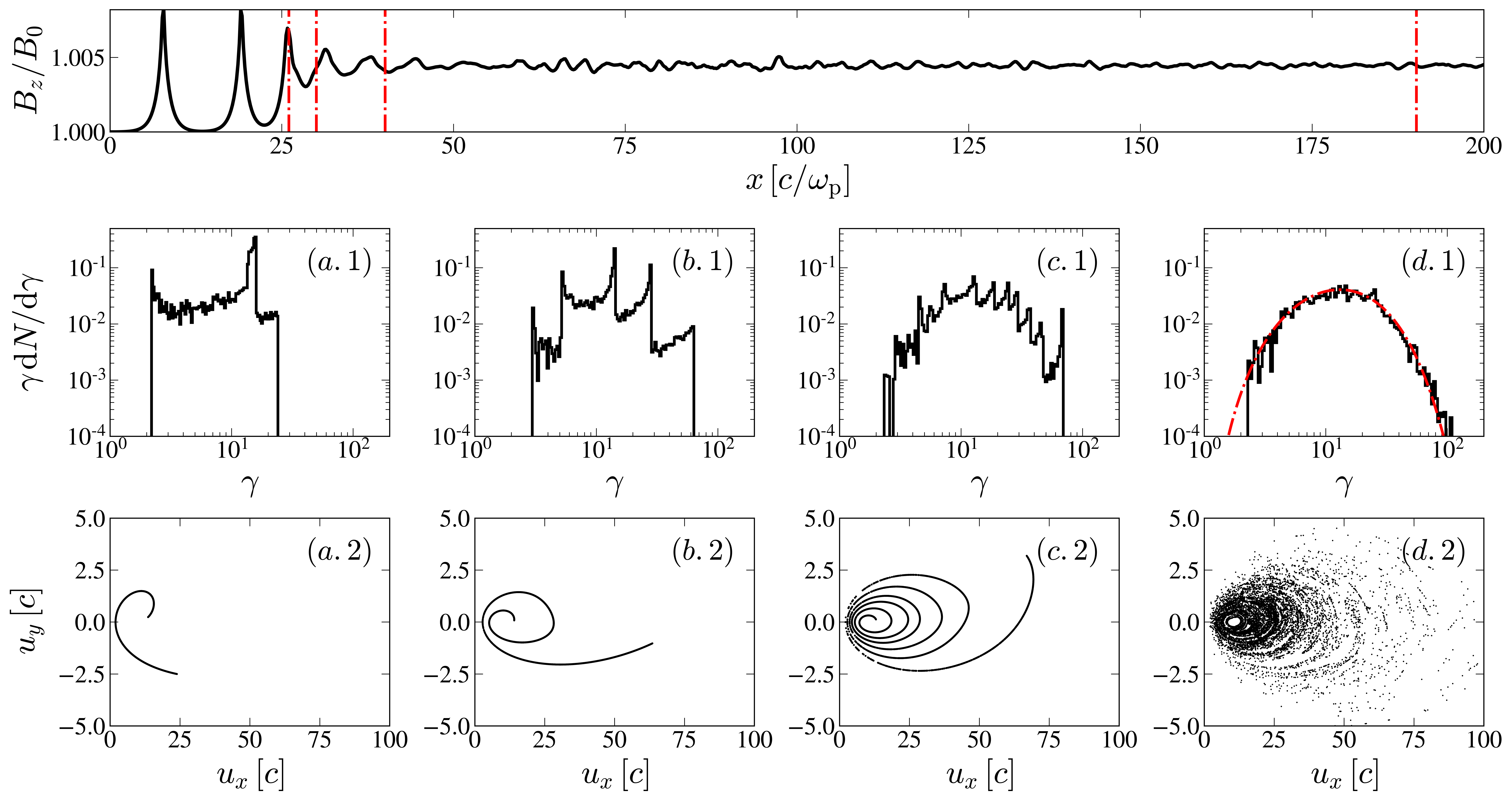}
  \caption{Phase-space profile of the beam velocity distribution for $N = 10^4$, $u_0 = 40$, and $\Ma = 4$. The top panel shows the magnetic-field compression ratio profile in the soliton frame. The dot-dashed red lines indicate the positions at which panels (a-d) are drawn. Panels (a-d.1) show the spectral evolution and convergence towards a 2D J\"uttner-Synge distribution, represented by a red dot-dashed line in panel (d.1). The corresponding momentum space distributions are displayed in panels (a-d.2),
  providing evidence for phase mixing between the different beam trajectories.}
  \label{fig:thermal}
\end{figure*}

\subsection{Chaos, Entropy, and Thermalization}\label{sec 2.2}

The profiles obtained thus far clearly exhibit stochasticity. To better quantify its nature, we now properly discuss the definition of emerging chaos. The first step in quantifying the erratic or chaotic behavior of the system's dynamics is derived from the spectrum of Lyapunov exponents:
\begin{align}
\mathbf{\Lambda} \,=\,\left\{\Lambda_i \,|\, i\,=\,-N,\dotsc,0,\dotsc,N \right\} \,. 
\end{align}
It can be seen that the flow is invariant under coordinate-reversal associated with the involution $R(u_x^i,\,u_y^i,\,B_z)\,=\,(u_x^i,\,-u_y^i,\,B_z)$. This implies that if $u_x^i(x), u_y^i(x), B_z(x)$ is a solution, so is $u_x(-x), -u_y(-x), B_z(-x)$. Therefore, growing trajectories are accompanied by decaying trajectories. We also identify two integrals of motion analogous to the conservation of momentum and energy fluxes discussed in~\cite{Alsop_1988}:
\begin{align}
&\partial_{\tilde{x}} \left( \frac{\tilde{B}}{\Ma^2} + \frac{1}{N}\, \sum^N_i \frac{\gamma^i}{u_0^2}\right)\,=\,0 \,,\\
&\partial_{\tilde{x}} \left( \frac{1}{2} \frac{\tilde{B}^2}{\Ma^2} + \frac{1}{N}\,\sum^N_i \tilde{u}^i_x\right)\,=\,0\,.
\end{align}
When interpreted as a dynamical system, these conserved quantities, along with the reversibility property, do not ensure that the system is conservative. However, we expect the spectrum to be symmetrically distributed around zero due to reversibility, with three vanishing Lyapunov exponents associated with this symmetry and the two integrals of motion. In such a large and unstable system, deriving $\Lambda_i$ is non-trivial due to contamination of the estimate by the largest exponents. To mitigate this issue, we approximate the spectrum from the mean growth rate of volumes in the tangent space of the set of $2 N+1$ equations~\eqref{eq:B_Nfluid} with continuous orthonormalization of the basis of the linearly independent vectors $\Phi\,=\,\left\{u_1,\,\dotsc,\,u_{2N+1}\right\}$~\cite{Benettin_1980}:
\begin{align}
    \partial_x\Phi\,=\,\mathcal{J}\cdot\,\Phi \,,\qquad{\rm with}\qquad \Phi(0)\,=\,I\,,
\end{align}
where $I$ is the identity matrix and $\mathcal{J}$ is the $(2 N + 1)\times (2 N + 1)$ Jacobian matrix of the system. The computational cost of the integration is greatly reduced due to the sparsity of the Jacobian matrix with only $8 N$ non-vanishing terms, allowing linear scaling of the integration with the number of fluids, though ultimately limited by the complexity of the continuous Gram–Schmidt orthogonalization.  

Figure~\ref{fig:Lyapunov}(a) shows the spectrum of $\Lambda_i$ for various $N\,=\,2,\,\dotsc,\,20$. In each case, the Lyapunov spectrum includes 3 vanishing values among the $2N+1$ characteristic exponents and $N-1$ positive exponents. The values at $|i|\,=\,1$ in Fig.~\ref{fig:Lyapunov} are degenerate. The spectrum shows a fast convergence of the maximum $\Lambda_i$ with the number of fluids. The lower boundary is set to $N\,=\,2$ for the simple reason that no chaotic behavior can emerge in systems of dimension lower than $D\,=\,2N+1 <5$ -- \emph{i.e.}, $N<2$ -- with two integrals of motion. As we have seen in Sec.~\ref{sec 1.1}, the one-fluid limit is indeed stable. The Lyapunov exponents also exhibit a dependence on $\Ma$, as shown in Fig.~\ref{fig:Lyapunov}(b). For $u_0\,=\,100$, a transition to chaos is observed around $\Ma\,\sim\,2$. For clarity, we use a low number of beams with $N\,=\,5$. According to Fig.~\ref{fig:Lyapunov}(a), it is sufficient to ensure reasonable convergence of the maximum value of $\Lambda_i$. The spectrum of Lyapunov exponents vanishes for $\Ma\,\lesssim\,2$. This indicates a phase transition. In the regime $\Ma\,\lesssim\,2$, stochastic effects become negligible, and we expect long-range soliton structures to emerge at low $\Ma$. The shock transition will significantly depart from the usual picture; their kinetic description and the associated emission will be discussed in a subsequent paper.

A salient feature of a shock is the net production of entropy. In this model, the production of information or entropy directly results from the sensitive dependence on initial conditions, where two infinitesimally close states become distinguishable over time. The specific Kolmogorov-Sinai entropy, normalized to the number of fluids, directly relates the rate of entropy generation to the Lyapunov exponents through the sum of the positive values in the spectrum: $s\,=\, \frac{k_{\rm B}}{\me N} \,\sum_{\Lambda_i>0} c\Lambda_i$~\cite{Eckmann_1985}. While a specific value of $s$ is of little interest here, the qualitative result provides an insightful perspective on the mechanism at the origin of entropy creation across the shock transition, and this, independently of the initial distribution. Additionally, $\mathbf{\Lambda}$, which converges rapidly with the number of fluids, also serves as an estimate for the inverse of the thermalization scale.

In the paradigm used throughout this paper, the thermalization of the plasma should be interpreted as an emergent property of the long-term dynamics of a non-integrable many-body system. A deeper understanding of the long-term effective thermalization of the set of fluid beams goes far beyond the scope of this paper. However, we here discuss the main features. Figure~\ref{fig:thermal} illustrates the evolution of the energy distribution and the phase-space profiles at different positions along a shock transition with $u_0\,=\,40$ and $\sigma\,=\,100$, for a system of $10^4$ fluids. Panels (a-d.1) in Fig~\ref{fig:thermal} demonstrate the progressive phase mixing between particle trajectories. Each panel corresponds to the instantaneous phase space distribution of beam velocities. 
The convergence towards a coarse-grained 2D Jüttner-Synge distribution is clear in panel \ref{fig:thermal}(d.2). We also note that thermalization is weakly dependent on the initial fluids spread. We tested this by using different initial conditions. The initial noise in the beam velocities associated with Fig.~\ref{fig:thermal} was set according to a uniform distribution. The same thermal profile follows from an initial distribution of velocities drawn at random from a boosted Jüttner-Synge distribution. 

A simple qualitative understanding of the thermalization process can be derived from the downstream profile of the magnetic field. Equation~\eqref{eq:B_Nfluid} indicates that phase mixing between a large number of weakly correlated trajectories results in a nearly constant compressed magnetic field. Small-amplitude fluctuations of the magnetic field provide a source of scattering. Additionally, other downstream modes, arising from plasma instabilities which are not accounted for in our analysis, will also likely contribute. Thermalization of the velocity components of the dynamical system should follow from effective small-angle scattering on the fluctuating magnetosonic modes propagating in the compressed downstream background field.

\section{Conclusion}\label{sec 3}

We have developed a fundamental  model for the structure of relativistically magnetized shock waves in  pair plasma.  A remarkable feature 
of this model is that shock dissipation does not require collective plasma interactions, rather it stems from nonlinear dynamics of 
charged particles in a stationary, quasiperiodic system of collisionless fast-magnetosonic solitons.
We find that particle trajectories in this highly nonlinear structure are unstable over a scale comparable to the distance between adjacent solitons, 
set by a characteristic wavenumber $k\,\sim\,\wpp/c$ that depends weakly on the upstream Alfv\'en Mach number $\Ma$ in the super-fast regime, $\Ma >1$. 
It is worth noting that in the downstream frame of the shock,  the weak dependence of $\lambda$ on $M_A$  readily implies that the wavenumber of the
transformed solitary structure scales 
as $ck_{\rm |d}\,\sim\,\gamma_{\rm sh|d}\,\wpp\,\sim\,\sqrt{\sigma}\,\wpp$, since $\gamma_{\rm sh|d}\,\sim\,\sqrt{\sigma}$ in the strongly magnetized regime~\cite{Kennel_1984}.  This has important implications for the generation of high-frequency emission in the shock, which will be 
discussed in the second paper of this series.

Using a reduced multi-fluid model that incorporates a self-consistent coupling of the fluids to the magnetic field, we have been able to capture the shock structure 
on kinetic scales and recover the shock jump conditions. Our analysis indicates that the mechanism underlying the shock transition is chaos in orbital dynamics. The chaotic nature of the system has been quantified by the set of Lyapunov exponents, and from their positive values we defined the rate of entropy creation associated with phase mixing. Finally, we revealed the thermalization of the downstream velocity distribution, interpreted as a small-angle scattering of particles in the fluctuating compressed magnetic field.

Our model provides a deeper insight into the underlying physics of relativistically magnetized shock waves. The shocks are of importance in many strongly magnetized astrophysical systems, including magnetars, pulsars, and blazars. While restricted to the regime $\sqrt{\sigma} < u_0\,\ll\,\sigma$, the present study directly opens promising avenues for the modeling of various classes of magnetized shocks, mediated or not by soliton structures. 

In this work, we have shown that dissipation, thermalization, entropy creation, and magnetic field compression across the shock transition all stem from the onset of chaos over a large range of parameters. However, soliton-driven instabilities, as well as their nonlinear coupling to the shock structure, may affect the shock dynamics. These aspects are also important for characterizing their observational signature and will be the topics of subsequent studies in this series on relativistically magnetized shocks.

\begin{acknowledgments}
\emph{Acknowledgments} We thank the anonymous referees for their useful comments that helped improve the manuscript. AV acknowledges financial support from the program AAP Tremplin FSI 2024. This work was granted access to the HPC resources of TGCC/CCRT under the allocation A0160415130 made by GENCI. AL wishes to thank the Center for Computational Astrophysics at the Flatiron Institute for their warm hospitality and support. This work was supported by a grant from the Simons Foundation (00001470, AL). This research was supported by the Munich Institute for Astro-, Particle, and BioPhysics (MIAPbP) which is funded by the Deutsche Forschungsgemeinschaft (DFG, German Research Foundation) under Germany´s Excellence Strategy – EXC-2094 – 390783311. 
\end{acknowledgments}

\bibliography{apssamp}

\end{document}